\documentclass[prl,aps,twocolumn]{revtex4}

\usepackage{amsmath,graphicx,epsfig}
\begin{document}

\def\pd#1#2{\frac{\partial #1}{\partial #2}}

\title{\bf Solitary waves on vortex lines in Ginzburg--Landau models}
\author {Natalia G. Berloff}
\affiliation {Department of Applied Mathematics and Theoretical Physics,
University of Cambridge, Wilberforce Road, Cambridge, CB3 0WA
}
\date {June 25, 2004}

\begin {abstract} Axisymmetric disturbances that preserve their form
  as they move along the vortex lines in  uniform Bose-Einstein
  condensates are obtained numerically by the solution of the
  Gross-Pitaevskii equation. A continuous family of such solitary
  waves is shown in the momentum ($p$) -- substitution energy ($\widehat{\cal E}$) plane
  with $p\rightarrow 0.09 \rho \kappa^3/c^2, \widehat{\cal E}\rightarrow 0.091
  \rho \kappa^3/c$ as $U \rightarrow c$, where $\rho$ is the density, $c$ is the speed of
  sound, $\kappa$ is the quantum of circulation and $U$ is the
  solitary wave velocity. It is shown that  collapse of a bubble
  captured by a vortex line leads to  the generation of such solitary
  waves in condensates. The various stages of  collapse are
  elucidated. In particularly, it is shown that during collapse
  the vortex core becomes significantly compressed and after collapse
  two solitary wave trains moving in opposite directions are formed on the
  vortex line.

  \end{abstract}
\pacs{ 03.75.Lm, 05.45.-a,  67.40.Vs, 67.57.De }
\maketitle
An important role in the dynamics of nonlinear systems is played by
solitary waves -- the localised disturbances of the uniform field that are
form-preserving and move with a constant velocity. They appear in
diverse contexts of science and engineering, such as fluid dynamics,
transport along macromolecules, fibre optic communications just to
name few. Considerable
interest attaches to determining the entire sequence of solitary waves
as they define
possible states that can be excited in the system. Understanding  the
production, motion and interactions of such solitary waves is one of the
most significant questions in nonlinear science. In
condensed matter systems  solitary waves are topological objects
since they owe their existence and perseverance to the topology of the
order parameter field describing a medium with a broken symmetry. In
this Letter I establish and study the production of a new class of solitary waves that move along
vortex lines/topological defects in conservative Ginzburg--Landau
systems. The discussion will be restricted to  condensed
matter system such as atomic Bose--Einstein condensates (BEC) where the evolution equation is the  Gross --
Pitaevskii (GP) model \cite{gp}; see equation (\ref{gp}) below (also known as defocusing nonlinear
Schr\"odinger equation in nonlinear optics). The
applications are not restricted to the condensed matter systems due to the generality of the
Ginzburg--Landau systems with implications  to the motion of
excitations along  cosmic strings  in the early Universe
\cite{srivastava} and  along topological defects in other ordered media:
liquid crystals \cite{chuang}, non--equilibrium patterns, etc. Finally, the new
solutions to the nonlinear Schr\"odinger equation which this Letter
presents are of interest  when so few have been
derived in multidimensions.

Symmetry-breaking transitions in equilibrium systems can be described
by an energy functional, ${\cal E} = \int {\cal L}\,dV,$ in terms
of a Lagrangian ${\cal L}$. The simplest form of the Lagrangian,
capable of describing the quenching beyond the critical point where
the disordered state becomes unstable and the symmetry is
spontaneously broken, is a Ginzburg-Landau Lagrangian ${\cal
  L}=\frac{1}{2}[|\nabla \psi|^2 + \frac{1}{2} (1 - |\psi|^2)^2]$ dependent on a
complex scalar order parameter $\psi.$ The  evolution equation describing the
relaxation to the equilibrium state in an energy--preserving
(conservative) system is given by the Euler--Lagrange
equation $\psi_t = -{\rm i}\partial {\cal E}/\partial \psi^*$ that we write
as
\begin{equation}
-2{\rm i} \pd \psi t =  \nabla^2 \psi +(1- |\psi|^2) \psi.
\label{gp}
\end{equation}
A phase singularity of a complex field $\psi$ given by $\psi=0$ is
called a quantised vortex or topological defect depending on a
particular application. The total change of phase around any closed
contour must be a multiple of $2 \pi$ and only quantised vortices with the
total change of phase $2\pi $ are topologically stable.

Such a general reasoning gives a simple explanation why equation (\ref{gp}) has  a universal meaning and has been
applied to a variety of systems \cite{pismen}. In particular, it described
accurately both equilibrium and dynamical properties of BEC
\cite{bec1}. The GP model has been remarkably
successful in predicting the condensate shape in an external
potential, the dynamics of the expanding condensate cloud and  the motion of
quantised vortices; it is also a popular qualitative model of
superfluid helium. For these systems equation (\ref{gp}) is written 
in dimensionless variables
such that the unit of length corresponds to the healing
length $\xi$, the speed of sound  $c=1/\sqrt{2}$,  and the density at
infinity  $\rho_\infty=|\psi_\infty|^2=1$. 

The straight-line vortex positioned along the $z-$ axis
  is obtained by rewriting
  (\ref{gp}) in cylindrical coordinates $(s,\theta, z)$ and using the ansatz 
  $\psi_0=R(s)\exp(i\theta)$. The resulting solution for $R(s)$ was
  found numerically in the first reference of \cite{gp} and approximated in
  \cite{pade}. The infinitesimal perturbations of a rectilinear vortex in
the GP model may be bound or free, depending on their angular and axial
wavenumbers, $m$ and $k$. The free waves radiate energy
acoustically to infinity, while the bound states do not. The
low frequency modes $m=1$, that displace the axis of the vortex, are
found to be 
bound for all $k$ \cite{pit61,robertsPRSA}. The low frequency modes
$m=2$ are also bound when $k$ is sufficiently large, but are free for
small $k$ \cite{robertsPRSA}.

A family of  fully three-dimensional
  solitary waves was found by Jones
and Roberts \cite{jr} who integrated  (\ref{gp}) numerically  and determined the entire sequence
of solitary wave solutions of the GP equation, such as vortex rings and finite amplitude sound waves named rarefaction
pulses. They showed  the location of the sequence on the
momentum, $p$,  energy, ${\cal E}$, plane, that I shall refer to as the
JR dispersion curve. In three dimensions they found two branches meeting at a cusp where $p$ and ${\cal E}$
assume their minimum values, $p_m$ and ${\cal E}_m$. As $p \to \infty$
on each branch, ${\cal E} \to \infty$. On the lower branch the
solutions are  asymptotic to  large vortex rings.                            
As ${\cal E}$ and $p$ decrease from infinity along the
lower branch, the solutions begin to lose their similarity to
large vortex rings.
Eventually, for a momentum $p_0$ slightly greater than $p_m$,
they lose their vorticity ($\psi$  loses its zero), and
thereafter the solitary solutions may better be described as
`rarefaction waves'. The upper branch consists entirely of these
and, as $p \to \infty$ on this branch, the solutions 
approach asymptotically the rational soliton solution of the Kadomtsev-Petviashvili Type I
 equation.

In what follows I determine the entire family of axisymmetric ($m=0$) solitary wave solutions
that move along the straight-line vortex and relate them to the JR
dispersion curve.
In equation (\ref{gp}) written in cylindrical coordinates
$(s,\theta,z)$ I take the ansatz $\psi = (R(s) + \phi(s,z)) \exp(i\theta)$,
 and assume that the
disturbance $\phi(s,z)$ moves with velocity $U$ in the positive
$z-$direction. In the frame of reference moving with the solitary
wave, $\phi(s,z)$ satisfies
\begin{eqnarray}
2 {\rm i} U \pd \phi z &=& \frac{1}{s}\frac{\partial}{\partial s}\biggl[s\pd \phi s\biggr]
+ \frac{\partial^2 \phi}{\partial z^2} - \frac{\phi}{s^2} \nonumber \\
&+&
(1-2R^2-R(\phi+2\phi^*)-|\phi|^2)\phi-R^2\phi^*.
\label{ugp}
\end{eqnarray}
The
disturbance is localised, so the boundary condition is $\phi(s,z)
\rightarrow 0$, as $|{\bf x}|\rightarrow \infty$ in all directions of
${\bf x}$. In view of the asymptotic expansions at infinity \cite{jr}, I
introduce stretched variables $z'=z$ and $s'=s\sqrt{1-2 U^2}$ and
 map the infinite domain onto the box
$(0,\frac{\pi}{2})\times(-\frac{\pi}{2},\frac{\pi}{2})$ using the transformation
$
\widehat z=\tan^{-1}(L z')
$ and $
\widehat s=\tan^{-1}(L s'),$
 where  $L$ is
a constant $\sim 0.1-0.4$.
Transformed equation (\ref{ugp}) was expressed in second-order finite
difference form using $250$x$200$ grid points, and the resulting nonlinear equations were solved by
Newton-Raphson iteration procedure using banded matrix 
 linear solver based on bi-conjugate gradient stabilised iterative method with
preconditioning. For each solitary wave  two quantities were
calculated: the  nonzero ($z-$) component of the
momentum \cite{jr} 
\begin{equation}
p=\frac{1}{2 {\rm i}}\int\biggl[\nabla \psi (\psi^*
  -1)-\nabla \psi^* (\psi-1)\biggr]dV
\label{p}
\end{equation}
 and the substitution energy,
$\widehat{\cal E},$ which is the difference between the energy of the 
vortex-solitary wave complex and the energy of the vortex line,
\begin{equation}
\widehat{\cal E}=\tfrac{1}{2}\!\int\!|\nabla \psi|^2\!-\!|\nabla\psi_0|^2 + \tfrac{1}{2}(1\! -\!
|\psi|^2)^2\!-\!\tfrac{1}{2}(1\!-\!|\psi_0|^2)^2\,dV.
\label{E}
\end{equation} 
 By performing the variation
$\psi\rightarrow\psi + \delta \psi$ in (\ref{p})--(\ref{E}) and discarding surface
integrals that vanish provided $\delta \psi \rightarrow 0$ for ${\bf
  x} \rightarrow \infty$, we see that $U=\partial \widehat{\cal E}/\partial p$,
  where the derivative is taken along the solitary wave sequence. The same
  expression is obeyed by the sequences of classical vortex rings in
  an incompressible fluid and by the solitary waves of \cite{jr}. I also
  note that if we multiply $\widehat{\cal E}$  by $z\partial
  \psi^*/\partial z$ and integrate by parts we get $\widehat{\cal
    E}=\int|\partial \psi/\partial z|^2 dV,$ similarly to the
  expression for the
  energy of the JR solitons \cite{jrp}. This expression was used as a
  check of the numerical accuracy. As $p \to \infty$,
 $\widehat{\cal E} \to \infty$ and  solitary wave solutions are
  represented by  large vortex rings moving along the vortex line.
  As $\widehat{\cal E}$ and $p$ decrease from infinity, the radii of the rings
  decrease and  for a momentum $p_0 \lesssim 78$,
$\phi(s,z)=0$ on $z-$axis only. To distinguish these solutions from
  vortex rings and to emphasise the analogy with the JR solitary
  waves, these solutions will be called rarefaction waves as well. Table 1 shows the
velocity, substitution
  energy, momentum and radius of the solitary wave solutions found.
  Figure 1 shows the momentum-energy curve of the solutions in
  comparison with the JR dispersion curve.
Notice that unlike the JR dispersion curve, there is no cusp on the
energy--momentum plane.
As $U\rightarrow c$ neither
$\widehat{\cal E}$ nor $p$ go to infinity, instead
$\widehat{\cal E} \rightarrow 32$ and $p\rightarrow 45$ which lies
below the JR cusp. The substitution energy, $\widehat{\cal E}$,  and
momentum, $p$, of our
vortex rings are larger that corresponding values of ${\cal E}$ and
$p$ of the JR rings moving with the same velocity. If the vortex rings of the
same radii are compared, our rings have lower energy and
momentum. Figure 2 shows the density isoplots, the density contour plots,
and the velocity vector fields of two representative solutions such as
a rarefaction pulse and a vortex ring. Notice  the vortex core
expansion at the centre of the vortex ring  due to a
decrease in pressure in high velocity region.

\bigskip

\noindent {\bf Table 1.} {\footnotesize The
velocity, $U$, substitution
  energy, $\widehat{\cal E}$, momentum, $p$ and radius, $b$, of the solitary wave solutions  moving along the
  straight-line vortex.}

\smallskip
\begin{tabular}{c | c c c c c c}
\hline
$U$& 0.4 & 0.45 & 0.5 & 0.55 & 0.6 & 0.61 \\
\hline
$\widehat{\cal E}$ & 142 & 113 & 90.7 & 72.4 & 56.9 & 54.0  \\ 
$p$ & 262 & 193 & 145 & 110 & 83.2 & 78.4  \\
$b$ & 4.18 & 3.62 & 3.08 & 2.41 & 1.05 & 0.1 \\
\hline
\end{tabular}

\smallskip
\begin{tabular}{c | c c c c c c}
\hline
$U$&  0.63 & 0.65 & 0.67 & 0.69 & 0.7 & 0.75 \\
\hline
$\widehat{\cal E}$ & 48.4 & 43.0 & 37.8 & 33.3 & 32.2 & 32.1 \\ 
$p$ &  69.4 & 61.0 & 53.1 & 46.5 & 45.0 & 45.0 \\
\hline
\end{tabular}

\smallskip
\begin{figure}[t]
\caption{(color online) The  dispersion curves for two families of the axisymmetric
  solitary wave solutions. The dashed line represents the JR
  dispersion curve. The part of the curve that corresponds to the
  vortex rings is shown in grey (red). The solid line gives the
  substitution energy as a function of momentum for the solitary waves
  moving along the vortex line with vortex rings shown in light grey (green).
}
\centering
\bigskip
\epsfig{figure=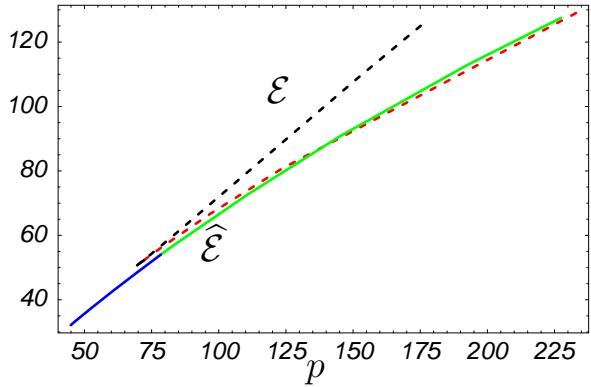, height = 1.9 in}
\begin{picture}(0,0)(0,0)
\put(-110,-4) {\Large $p$}
\put(-125,100) {\Large ${\cal E}$}
\put(-150,40) {\Large $\widehat{\cal E}$}
\end{picture}
\label{cusp}
\end{figure}
\begin{figure}
\caption{(colour online) Two solitary wave solutions moving with
  velocity $U=0.45$ (left side) and $U=0.69$ (right side) along the
  straight-line vortex. The density isoplots at $\rho=0.1$ (left) and
  $\rho=0.3$ (right) are shown on two top panels. The density contour 
  plots ($0.1,0.3, 0.5, 0.7, 0.9$) at the cross-section $\theta=0$ are shown on two middle panels. The
  velocity fields with contour plots ($0.05,0.1,0.3,0.5$) of the velocity magnitude,
  $\sqrt{v_s^2+v_z^2}$, are given at the bottom.
}
\centering
\epsfig{figure=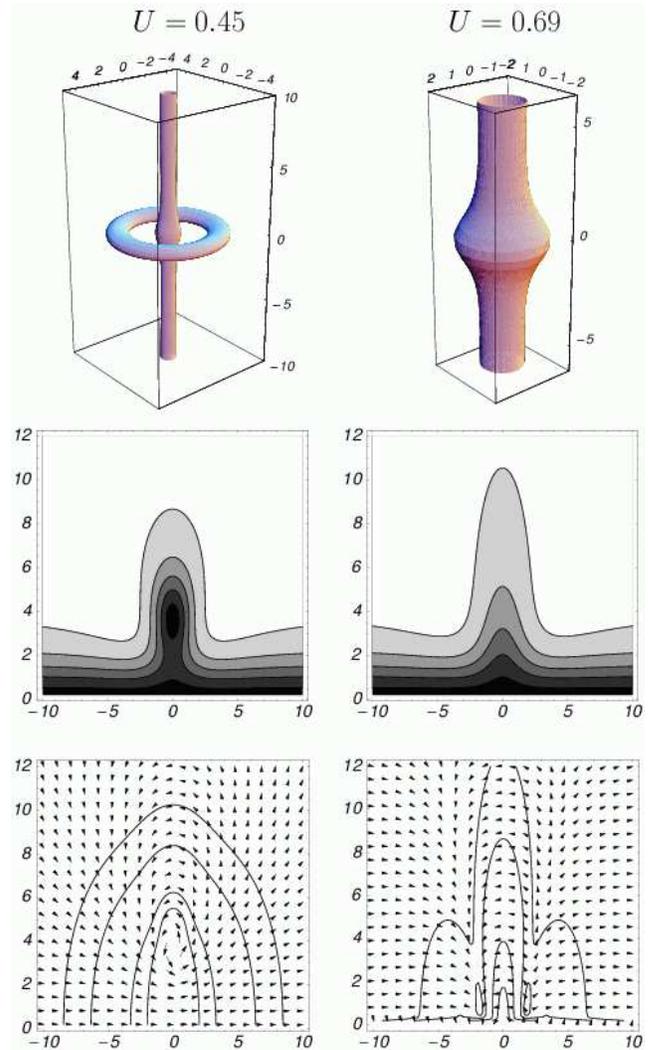, height = 5.5in}
\end{figure}

A question that arises after new solutions are found theoretically is how to
create them in an actual physical system. In \cite{bubble} 
 we established a new mechanism of vortex nucleation by collapsing
 bubbles in the context of the GP model. These results referred to collapse of cavitated bubbles  generated by ultrasound in the megahertz 
frequency range that have been observed to produce quantised vortices in
superfluid helium \cite{finch}. Also, vortices form as a result
of bubbles colliding  during a first-order phase transition of an early Universe
\cite{srivastava}. In \cite{capture}  we have shown
that  a soft bubble, carved out in the surrounding fluid by an
electron  through its zero-point motion, becomes trapped in vortex
lines. The Bernoulli effect of the flow created by the flow
circulation  around the  vortex propels the bubble and vortex towards
one another with a force approximately proportional to $s^{-3}$, where
$s$ is the closest distance between them. As the bubble becomes
trapped in the vortex core, the flow round the bubble acquires
circulation that it previously could not posses. After the emission of
Kelvin waves, that were excited on the vortex core during the capture,
the bubble-vortex complex stabilises to an axisymmetric form depicted
in Figure 2 of \cite{capture}; see also Figure 3 ($t=0$) below. One
could expect that a similar capture of bubbles created by ultrasound
takes place in  BEC and that vacuum bubbles get trapped
in cosmic strings. The captured bubble will than collapse sending
axisymmetric waves along the vortex line. To elucidate the stages of
this collapse I performed the numerical simulations of the GP equation (\ref{gp})
starting with the initial condition
\begin{equation}
\psi({\bf x},t=0)=\begin{cases}R(s)e^{i\theta}{\rm
                    tanh}(\frac{r-a}{\sqrt{2}})& \textstyle{\rm if}
                    \quad  r>a \\
                    0        & \textstyle{\rm if} \quad  0\le r \le a,
       \end{cases}
\label{ic}
\end{equation}
where $r^2=x^2+y^2+z^2$.
 Initial state (\ref{ic})
 gives an accurate representation of the
stationary complex, that consists of the straight-line vortex at $x=0,
y=0$ and the
bubble of radius $a$ centred at the origin.
The surface of the bubble
is assumed to be an infinite potential barrier to the condensate
particles, so no bosons can be found inside the bubble
 ($\psi=0$) before the collapse, and this is why ${\rm
  tanh}(\ell/\sqrt{2})$, which is the wavefunction of the condensate
distance $\ell$ away from a solid wall, is relevant here.   `Softer' bubbles that allow some
condensate penetration were also considered by reducing the slope of
the hyperbolic tangent, but no significant difference was detected. I performed fully three-dimensional 
calculations 
for cavities of various radii in a 
computational cube with  sides of $200$ healing lengths
\cite{numer2}. Figure 3 shows  the density  plots of the portion of the cross-section at $y=0$
at  various times after the collapse of a bubble of radius
$a=10$. Figure 4 depicts density contours 
$\rho/\rho_\infty = 4/5$ at times $t=0, 20$ and $60$. The
 time-dependent evolution of the condensate during and after bubble's collapse   involves several stages.  During the first stage 
dispersive and nonlinear wave-trains are generated at the surface of
the  bubble.
 This stage of the
 evolution is characterised by a flux of particles towards the centre
 of the cavity. This creates an inward force acting on the vortex core
 reducing the cross-sectional area of the core. The
 reduction by a factor of $1.5$ is seen on $t=15$ snapshot of Figure 3 and on
the density contour of Figure 4 at $t=20$. The next stage in the
evolution is outward expansion of the condensate that overfilled the
cavity. The instability mechanism for
collapsing bubbles in the absence of the straight-line vortex, which we described in  detail in
\cite{bubble}, sets in,
leading to the production of vortex rings and rarefaction pulses mostly along the vortex line, as the energy and momentum
necessary for their creation is lower there. As the train of solitary
waves starts moving away from the collapsed bubble (see Figures 3 and 4), the distance
between them increases since they move with different velocities. In time
each individual solitary wave approaches its localised form found in
the first part of this paper; see the insert of Figure 4. During
 collapse of a bubble of a larger radius ($a \gtrsim 28$) vortex rings
 will be generated together with rarefaction pulses on the vortex
 line; see the insert of Figure 3.

\begin{figure}[t]
\caption{(colour online) Collapse of the bubble of radius $a=10$
  trapped by the vortex line. The density plots of the
  cross-section of the solution of (\ref{gp}) with initial state
  (\ref{ic}) at $y=0$, $x \in [-50,50]$, $z \in [0,100]$ are
  shown. The insert shows a vortex ring travelling along the
  $z-$axis generated after collapse of the bubble of radius
  $a=50$ at $t=100$. Two vortex rings of smaller radii are also seen
  on the left and on the right
  of it.
 Both low and high density regions are shown in
darker shades to emphasise  intermediate density regions.
}
\centering
\epsfig{figure=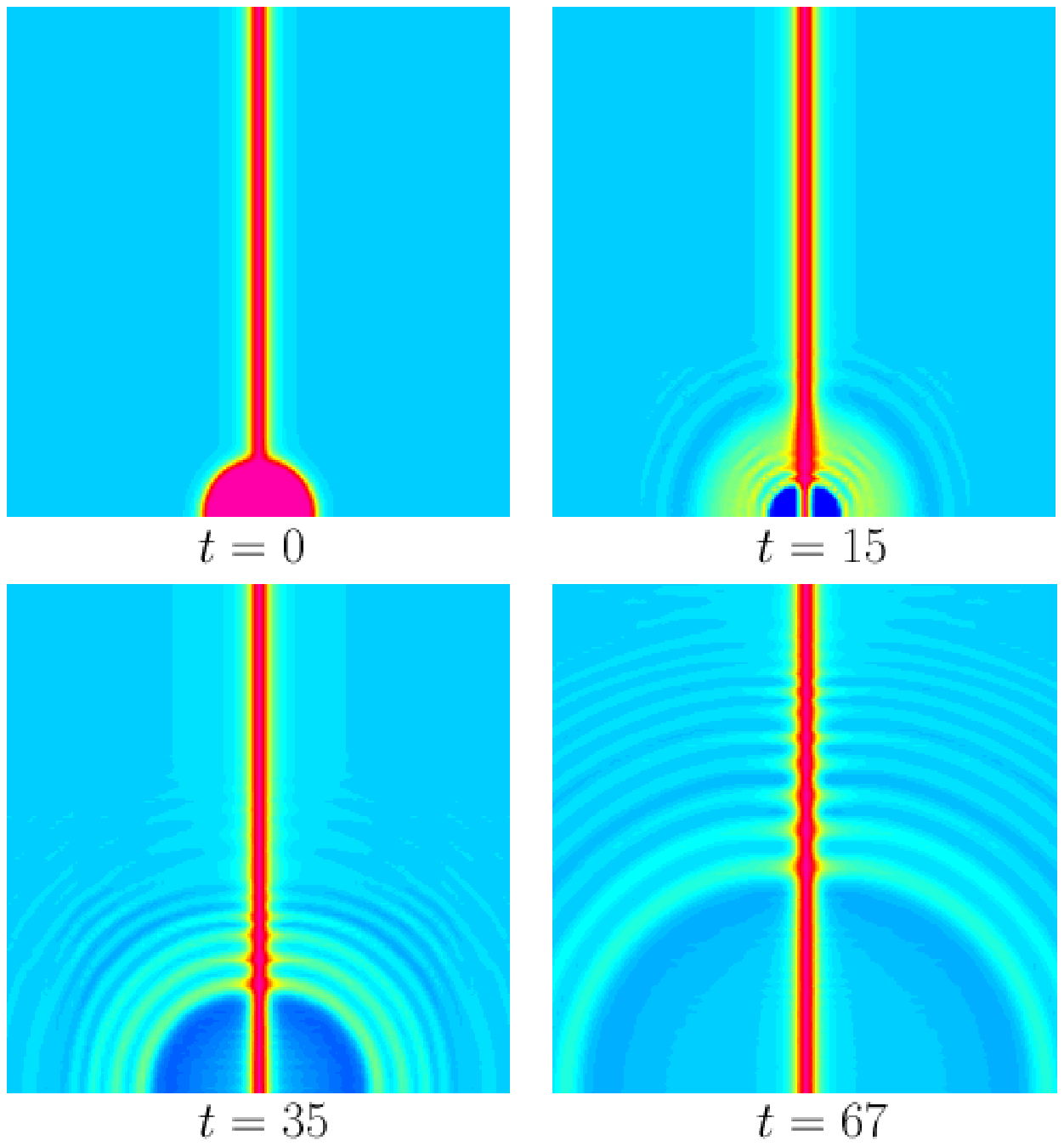, height = 3.4in}
\begin{picture}(0,0)
\put(-210,200) {\vector(1,0){20}}
\put(-210,200) {\vector(0,1){20}}
\put(-220,215) {$z$}
\put(-195,190) {$x$}
\put(-45,70) {\epsfig{figure=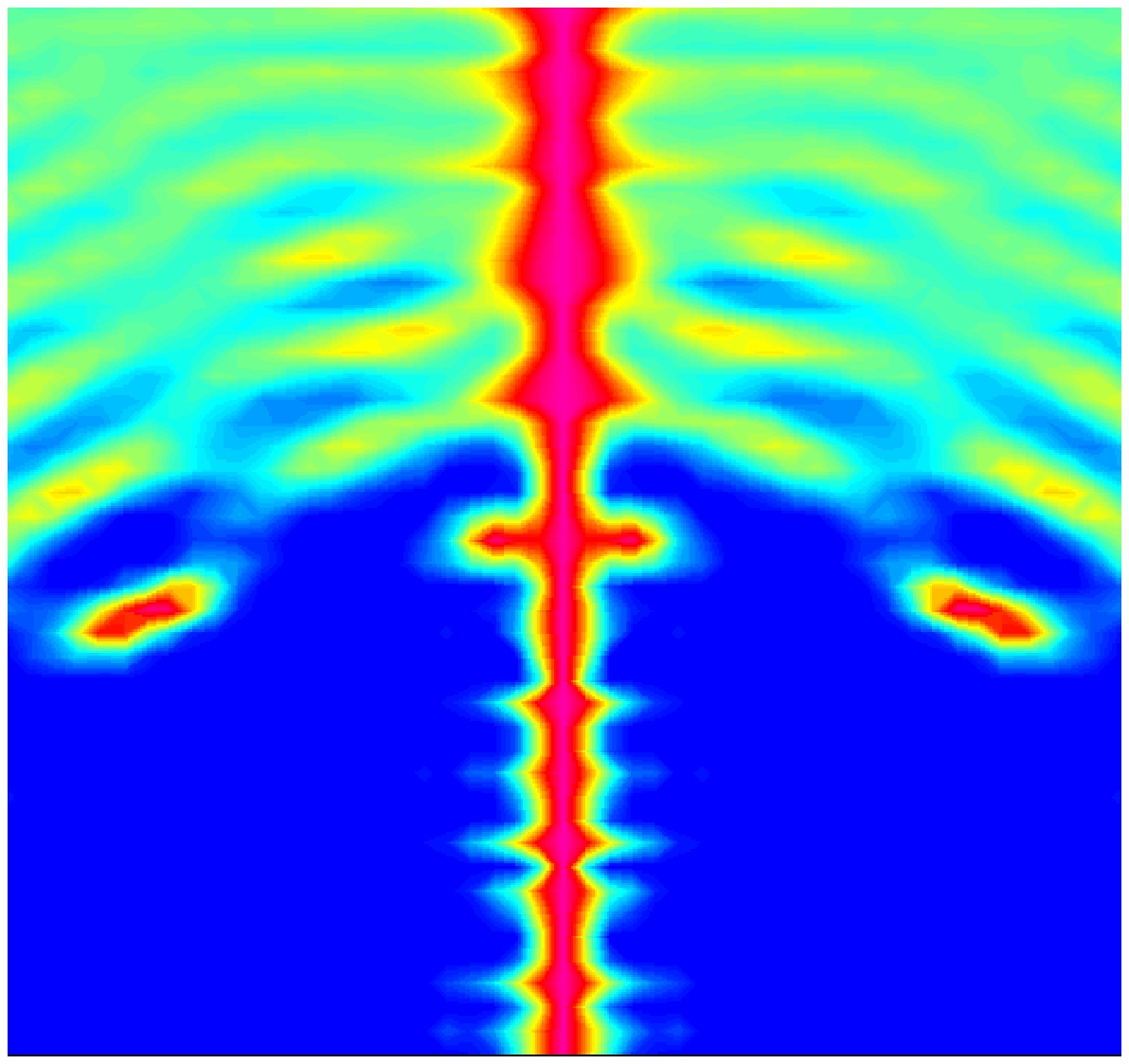, height = 0.5in}}
\end{picture}
\vskip -0.2 in
\end{figure}

\begin{figure}[t]
\caption{(colour online) Collapse of the bubble of radius $a=10$. The
  contour $\rho/\rho_\infty=4/5$ is shown at $t=0$ (green solid line), $t=20$ (blue dotted line),
  and $t=60$ (red dashed line). The density contour plot of the
  condensate  showing a well-separated solitary wave at $t=166$ is given
  in the insert (compare this with the middle right panel of Figure 2).
}
\centering
\epsfig{figure=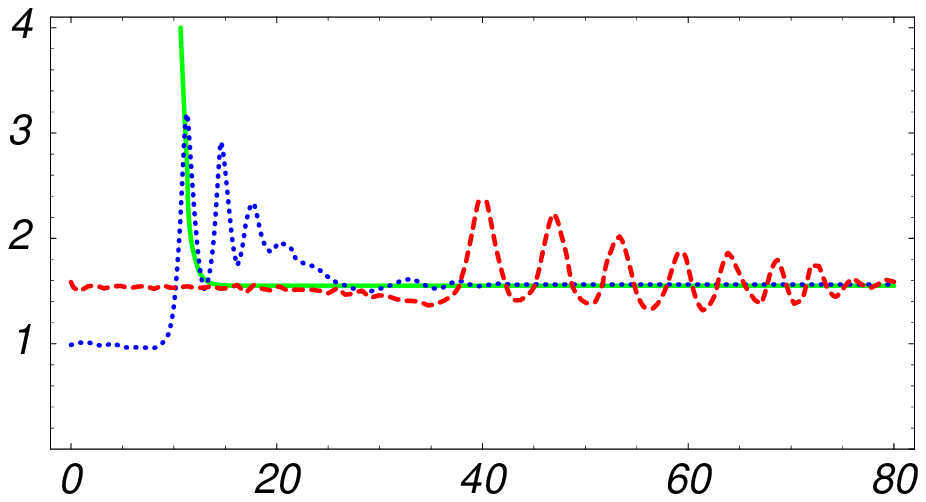, height = 1.8in}
\begin{picture}(0,0)
\put(60,75) {\epsfig{figure=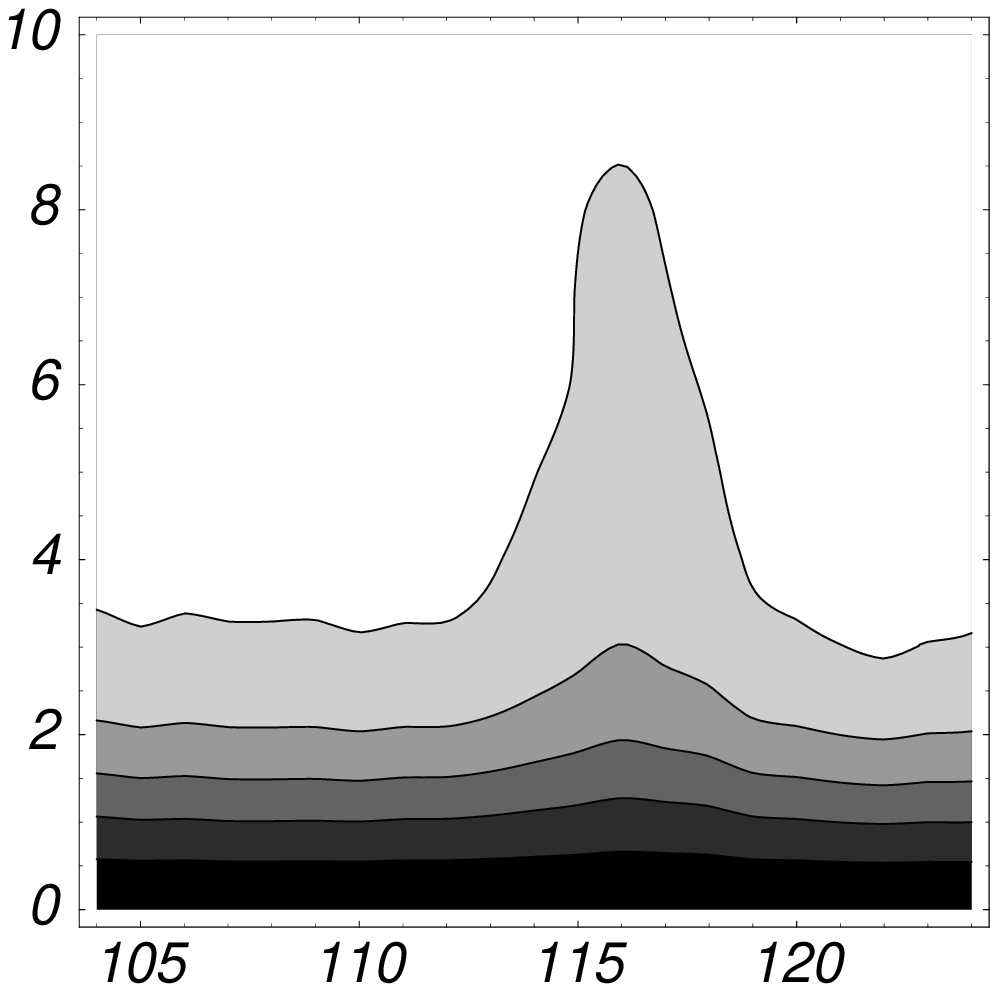, height = 0.8in}}
\put(110,30) {\Large $z$}
\put(-95,120) {\Large $x$}
\end{picture}
\vskip -0.2 in
\end{figure}

The author is grateful to Professors Paul Roberts and Boris Svistunov for useful comments
about this manuscript. The support from  NSF grant DMS-0104288 is
acknowledged. 


\begin{thebibliography}{99}
\bibitem{numer2}
I used the same numerical method as in 
work \cite{br7}. The faces of the computational box were
open to allow sound waves to escape.
\end{thebibliography}
\end{document}